
%
\magnification=1200
\overfullrule=0pt
\baselineskip=26 true bp

\def\gl{\mathrel{\raise1ex\hbox{$>$\kern-.75em\lower1ex\hbox{$<$}}}}
\def\lg{\mathrel{\raise1ex\hbox{$<$\kern-.75em\lower1ex\hbox{$>$}}}}
\def\gtwid{\mathrel{\raise.3ex\hbox{$>$\kern-.75em\lower1ex\hbox{$\sim$}}}}
\def\ltwid{\mathrel{\raise.3ex\hbox{$<$\kern-.75em\lower1ex\hbox{$\sim$}}}}
\def\sqr#1#2{{\vcenter{\hrule height.#2pt
      \hbox{\vrule width.#2pt height#1pt \kern#1pt
         \vrule width.#2pt}
      \hrule height.#2pt}}}

\def\eg{\hbox{{\it e.\ g.}}}\def\ie{\hbox{{\it i.\ e.}}}\def\etc{{\it etc.}}
\def\vs{{\it vs.}}\def\vv{{\it vice versa}}
\def\apr{{\it a priori}}


\def\leaderfill{\leaders\hbox to 1em{\hss.\hss}\hfill}

\def\ref#1{${}^{#1}$}
\newcount\eqnum \eqnum=0  
\newcount\eqnA\eqnA=0\newcount\eqnB\eqnB=0\newcount\eqnC\eqnC=0\newcount\eqnD\eqnD=0
\def\eqnoi{\global\advance\eqnum by 1\eqno(\the\eqnum)}
\def\eqnai{\global\advance\eqnum by 1\eqno(\the\eqnum{a})}
\def\eqnbi{\eqno(\the\eqnum{b})}
\def\eqnoA{\global\advance\eqnA by 1\eqno(A\the\eqnA)}
\def\eqnoB{\global\advance\eqnB by 1\eqno(B\the\eqnB)}
\def\eqnoC{\global\advance\eqnC by 1\eqno(C\the\eqnC)}
\def\eqnoD{\global\advance\eqnD by 1\eqno(D\the\eqnD)}
\def\back#1{{\advance\eqnum by-#1 Eq.~(\the\eqnum)}}
\def\backs#1{{\advance\eqnum by-#1 Eqs.~(\the\eqnum)}}
\def\backn#1{{\advance\eqnum by-#1 (\the\eqnum)}}
\def\backA#1{{\advance\eqnA by-#1 Eq.~(A\the\eqnA)}}
\def\backB#1{{\advance\eqnB by-#1 Eq.~(B\the\eqnB)}}
\def\backC#1{{\advance\eqnC by-#1 Eq.~(C\the\eqnC)}}
\def\backD#1{{\advance\eqnD by-#1 Eq.~(D\the\eqnD)}}
\def\last{{Eq.~(\the\eqnum)}}                  
\def\lasts{{Eqs.~(\the\eqnum)}}                
\def\lastn{{(\the\eqnum)}}                     
\def\lastA{{Eq.~(A\the\eqnA)}}\def\lastB{{Eq.~(B\the\eqnB)}}
\def\lastC{{Eq.~(C\the\eqnC)}}\def\lastD{{Eq.~(D\the\eqnD)}}
\newcount\refnum\refnum=0  
\def\refi{\smallskip\global\advance\refnum by 1\item{\the\refnum.}}

\newcount\rfignum\rfignum=0  
\def\rfigi{\medskip\global\advance\rfignum by 1\item{Figure \the\rfignum.}}

\newcount\fignum\fignum=0  
\def\figi{\global\advance\fignum by 1 Fig.~\the\fignum}

\newcount\rtabnum\rtabnum=0  
\def\rtabi{\medskip\global\advance\rtabnum by 1\item{Table \the\rtabnum.}}

\newcount\tabnum\tabnum=0  
\def\tabi{\global\advance\tabnum by 1 Table~\the\tabnum}

\newcount\secnum\secnum=0 
\def\chap#1{\global\advance\secnum by 1
\bigskip\centerline{\bf{\the\secnum}. #1}\smallskip\noindent}

\def\p2d#1#2{{\partial^2 #1\over\partial #2^2}} 
\def\t2d#1#2{{d^2 #1\over d #2^2}} 


\def\ith{{$i^{\rm th}$}}
\def\jth{{$j^{\rm th}$}}
\def\kth{{$k^{\rm th}$}}\def\2kth{{$2k^{\rm th}$}}

\def\n-th{{$(n-1)^{\rm th}$}}

\def\N-th{{$(N-1)^{\rm th}$}}

\def\0th{$0^{\rm th}$}
\def\1st{$1^{\rm st}$}
\def\2nd{$2^{\rm nd}$}
\def\3rd{$3^{\rm rd}$}
\def\4th{$4^{\rm th}$}
\def\5th{$5^{\rm th}$}
\def\5th{$6^{\rm th}$}
\def\6th{$7^{\rm th}$}
\def\7th{$7^{\rm th}$}
\def\8th{$8^{\rm th}$}
\def\9th{$9^{\rm th}$}


\def\G{\Gamma}
\def\d{\delta}\def\D{\Delta}

\def\l{\lambda}
\def\m{\mu}
\def\n{\nu}
\def\p{\pi}

\def\t{\tau}

\def\acp #1 #2 #3 {{\sl Adv.\ Chem.\ Phys.} {\bf #1}, #2 (#3)}
\def\cmp #1 #2 #3 {{\sl Commun.\ Math.\ Phys.} {\bf #1}, #2 (#3)}
\def\cp #1 #2 #3 {{\sl Chem.\ Phys.} {\bf #1}, #2 (#3)}
\def\eul #1 #2 #3 {{\sl Europhys.\ Lett.} {\bf #1}, #2 (#3)}
\def\jcp #1 #2 #3 {{\sl J.\ Chem.\ Phys.} {\bf #1}, #2 (#3)}
\def\jdep #1 #2 #3 {{\sl J.\ de Physique I} {\bf #1}, #2 (#3)}
\def\jdepl #1 #2 #3 {{\sl J. de Physique Lett.} {\bf #1}, #2 (#3)}
\def\jetp #1 #2 #3 {{\sl Sov.\ Phys.\ JETP} {\bf #1}, #2 (#3)}
\def\jetpl #1 #2 #3 {{\sl Sov. Phys.\ JETP Letters} {\bf #1}, #2 (#3)}
\def\jmp #1 #2 #3 {{\sl J. Math. Phys.} {\bf #1}, #2 (#3)}
\def\jpa #1 #2 #3 {{\sl J. Phys.\ A} {\bf #1}, #2 (#3)}
\def\jpchem #1 #2 #3 {{\sl J. Phys.\ Chem.} {\bf #1}, #2 (#3)}
\def\jpsj #1 #2 #3 {{\sl J. Phys.\ Soc. Jpn.} {\bf #1}, #2 (#3)}
\def\jsp #1 #2 #3 {{\sl J. Stat.\ Phys.} {\bf #1}, #2 (#3)}
\def\nat #1 #2 #3 {{\sl Nature} {\bf #1}, #2 (#3)}
\def\pA #1 #2 #3 {{\sl Physica A} {\bf #1}, #2 (#3)}
\def\pB #1 #2 #3 {{\sl Physica B} {\bf #1}, #2 (#3)}
\def\pD #1 #2 #3 {{\sl Physica D} {\bf #1}, #2 (#3)}
\def\pla #1 #2 #3 {{\sl Phys.\ Lett. A} {\bf #1}, #2 (#3)}
\def\plb #1 #2 #3 {{\sl Phys.\ Lett. B} {\bf #1}, #2 (#3)}
\def\pr #1 #2 #3 {{\sl Phys.\ Rev.} {\bf #1}, #2 (#3)}
\def\pra #1 #2 #3 {{\sl Phys.\ Rev.\ A} {\bf #1}, #2 (#3)}
\def\prb #1 #2 #3 {{\sl Phys.\ Rev.\ B} {\bf #1}, #2 (#3)}
\def\pre #1 #2 #3 {{\sl Phys.\ Rev.\ E} {\bf #1}, #2 (#3)}
\def\prl #1 #2 #3 {{\sl Phys.\ Rev.\ Lett.} {\bf #1}, #2 (#3)}
\def\prslA #1 #2 #3 {{\sl Proc.\ R.\ Soc.\ London, Ser.\ A} {\bf #1}, #2 (#3)}
\def\rmp #1 #2 #3 {{\sl Rev.\ Mod.\ Phys.} {\bf #1}, #2 (#3)}
\def\rpp #1 #2 #3 {{\sl Rep.\ Progr.\ Phys.} {\bf #1}, #2 (#3)}
\def\sci #1 #2 #3 {{\sl Science} {\bf #1}, #2 (#3)}
\def\sciam #1 #2 #3 {{\sl Scientific American} {\bf #1}, #2 (#3)}
\def\usp #1 #2 #3 {{\sl Sov.\ Phys.\ Usp.} {\bf #1}, #2 (#3)}
\def\zpb #1 #2 #3 {{\sl Z. Phys.\ B} {\bf #1}, #2 (#3)}
\def\zpc #1 #2 #3 {{\sl Z. Phys.\ Chem.} {\bf #1}, #2 (#3)}


\centerline{\bf Annihilation of Charged Particles}

\bigskip\bigskip
\centerline  {I.~Ispolatov and P.~L.~Krapivsky\footnote{$^\star$}
{Present address: Courant Institute of Mathematical Sciences,
New York University, New York, NY 10012}}
\bigskip
\centerline{\sl Center for Polymer Studies and Department of Physics}
\centerline{\sl Boston University, Boston, MA 02215}

\vskip 1in


\centerline{ABSTRACT}
{\smallskip\noindent
The kinetics of irreversible annihilation of charged
particles performing overdamped motion induced by long-range interaction
force, $F(r)\sim r^{-\l}$, is investigated.   The system exhibits rich
kinetic behaviors depending on the force exponent $\l$. In one
dimension we find that the densities decay as $t^{-1/(2+\l)}$
and $t^{-1/(1+2\l)}$ when $\l>1$ and $1/2<\l<1$, respectively,
with logarithmic correction at $\l=1$. For $\l \leq 1/2$,
the asymptotic behavior is shown to be dependent on system size.}
{
\narrower\narrower\bigskip\noindent
PACS number(s): 64.60.Cn, 64.60.My
}

\vfill\eject


\chap{INTRODUCTION}

The kinetics of two-species diffusion-controlled annihilation reaction,
$A+B\to 0$, between {\it uncharged} particles has been a subject of
extensive research for almost 20 years [1].  For sufficiently low
spatial dimension, $d<4$,  even under homogeneous initial conditions,
large-scale
heterogeneities arise that invalidate classical kinetic laws.  Much
less is known about annihilation reaction between {\it charged}
particles  with long-range power-law interaction, $F(r)\sim r^{-\l}$.
An important case of Coulomb interaction ($\l=d-1$ in $d$ dimensions)
has been treated in a few studies [2-5] for $d=2$ and 3.
However,  some of these works were based on unjustified approximations
while  others were based solely on numerical simulations, so their results
are also uncertain. (Note that even for annihilation of uncharged
particles in three dimensions  the asymptotic regime is hardly reached
on modern computers).   We, therefore, see that the Coulomb case still
deserves further investigation.  Other values of the interaction
exponent $\l$ also naturally appear in applications with particles
being dipoles, defects, vortices, monopoles, disclinations, \etc\
One important example is the quench of a one-dimensional
Ising system from a disordered state to an ordered state. If spins
interact via long-range potential [6], the Hamiltonian may be expressed
in terms of interacting domain walls [7]. There are two types of domain
walls in the system, the domain walls with "up" spins to the right and
"down" to the left
($A$-walls), and \vv\ ($B$-walls).  Thus, an alternating domain wall
sequence $...ABABAB...$ is formed.
Domain walls annihilate upon colliding, $A+B\to 0$, but since the
alternating structure persists in time, the reaction process is, in fact,
equivalent to the single-species annihilation, $D+D\to 0$.
This system has been recently investigated [7-9], and it was shown
that particle concentration decays as $t^{-1/(1+\l)}$.

In this paper we consider a truly two-species annihilation model
where the initial distribution of interacting particles ("charges")
is random (Poissonian).  The  forces between charges are
assumed to be proportional to $r^{-\l}$, with similar charges
repelling each other and dissimilar attracting each other.  Compared to the
single-species case, the two-species annihilation exhibits more rich
kinetic behavior including the dependence on the system size.  In this
study we focus on one-dimensional systems which allow us to find rather
convincing numerical support for our scaling predictions.

The paper has the following structure: In the next Section we
formally introduce the model, an ensemble of interacting particles
in one dimension with overdamped dynamics.
Then, relying on heuristic arguments, we obtain density decay
exponents for different values of $\l$. In Section III, we present the
results of numerical simulations. Finally, in Section IV we discuss  possible
generalizations of the model including higher dimensionality and
ballistic motion, and make general conclusions.

\bigskip\bigskip
\chap{THE MODEL AND SCALING ARGUMENTS}

We consider two-species systems
containing $A$- and $B$-type particles with charges $+1$
and $-1$ for "particles" and "antiparticles", respectively.
Particles of both species move continuously in one dimension and
interact via long-range force, $F=qq'/r^\l$ (for charges $q$ and $q'$
separated by the distance $r$).  Initially, $A$- and $B$-type particles
are randomly distributed with equal concentrations,
for simplicity we put them equal to 1.

Total force acting on the \ith\ particle is equal to a sum of
pairwise forces:
$$
F_i=q_i\sum_{j\not=i}{q_j (x_i-x_j)\over \vert x_i-x_j \vert ^{\l+1}},
\eqnoi
$$
where $q_k=\pm 1$ and $x_k$ are charge and coordinate of the \kth\
particle. We will ignore particle inertia; \ie, motion of
particles is assumed to be overdamped. Therefore, the velocity
of each particle $v_i$ is proportional to the total force $F_i$
acting on it, $v_i=\m F_i$. (In the following, we set the mobility $\m$
equal to 1).  We will also ignore particle diffusion; that is,
we will assume that the drift dominates the random walk effects.
When two dissimilar particles collide, both of them irreversibly disappear;
collisions between particles of the same species are impossible
because of repulsion.  To summarize, we consider two-species
annihilation of particles undergoing overdamped noiseless motion.
We will see that for a sufficiently large force exponent ($\l>4$ in
one dimension), the noise actually dominates the drift, and thus
well-known diffusion-controlled kinetic behavior [1] emerges.
However, for small $\l$, we expect that the long-time behavior is
correctly described by our noiseless model.  We will also briefly
discuss a model where the motion is ballistic (Sec.V).

Let us now consider time evolution of the system.  We cannot \apr\
expect that a mean-field description holds in low dimensions,
especially in one dimension.  Remember that the breakdown of
the mean-field behavior in reaction-diffusion models is generally
attributed to the formation of single-species domains [1]. We
assume that the same takes place in our model (at least in one
dimension, the formation of domains is inevitable).

Suppose that at time $t$ the  length of a typical
single-species domain is $L(t)$. It means that an average number of
particles in such a domain is equal to initial imbalance of majority and
minority species on the length $L(t)$, which for Poissonian initial
distribution with the density one is of the order of  $\sqrt{L(t)}$.
Therefore, the concentration $n(t)$ in a typical domain behaves as
$$
n(t) \sim 1/ \sqrt {L(t)}.
\eqnoi
$$

To get an insight on how a typical domain length changes in time, we
consider motion of a single particle $\tilde A$ on a domain edge
(Fig.~1). Here we assume that the system is entirely formed of
well-defined domains of typical length $L$.
The total force acting on $\tilde A$ from the particles on its left may
be evaluated as:
$$
F\sim \sum_{j=1}^{M}{1\over x_j^\l}-\sum_{j=1}^{M} {1\over {(x_j+L)}^\l}+
\sum_{j=1}^{M}{1\over {(x_j+2L)}^\l}- \cdots .
\eqnoi
$$
Here $M$ is a typical number of particles in a domain,
$M \sim \sqrt {L}$.  Each sum in the left-hand side of \last\
expresses a contribution to the net force from a particular domain;
to get the total force these contributions have been added.
It is clear that the force exerted on $\tilde A$ from
all the particles to the right may be calculated in exactly the same
way. To simplify the matter even further we assume that $x_j \simeq R
\times j$,
where $R$ is an average interparticle separation; $R=L/M$.  Rewriting
\last\ through $R$ and $M$ gives
$$
F\sim {1\over R^\l}\left[\sum_{j=1}^{M}{1\over j^\l}-\sum_{j=1}^{M}
{1\over {(j+M)}^\l}+
\sum_{j=1}^{M}{1\over {(j+2M)}^\l}- \cdots\right].
\eqnoi
$$
Depending on the value of the force exponent $\l$,
different situations appear. For $\l >1$, the first sum converges
to a finite value for $M \rightarrow \infty$ while the other sums
approach to zero as $M^{-(\l-1)}$ which means that only charges from
the left and right nearest neighbor domains essentially contribute
to the total force. For $0<\l< 1$, all sums diverge as $M^{1-\l}$
as $M \rightarrow \infty$. The total force, being the sum of
monotonically decreasing sign-alternating terms, is of order of the
contribution of the first domain. In the borderline case $\l=1$, the
first sum diverges as $\log M$ while the others terms are monotonically
decreasing, sign-alternating, and finite. The dominate contribution is
again provided by the nearest domains. To sum up,
$$
F \sim {1\over R^\l}\times \cases{M^{1-\l} & if \quad $0<\l<1$,\cr
\cr
\log{M} & if \quad $\l=1$, \cr
\cr
1 & if \quad $\l>1$. \cr}
\eqnoi
$$

On the other hand, a typical rate of change of the domain length is of
order of the velocity of any of its edges. Recalling that
$R \sim M\sim \sqrt{L}$, we obtain
$$
dL/dt \sim F \sim    \cases{L^{1/2-\l} & if \quad $0<\l<1$,\cr
\cr
L^{-1/2}\log{L} & if \quad $\l=1$, \cr
\cr
L^{-\l/2} & if \quad $\l>1$. \cr}
\eqnoi
$$

Solving (6) for $L(t)$ and using relation (2),
we finally write for the density decay asymptotics:
$$
n(t) \sim  \cases{t^{-{1\over 1+2\l}} & if \quad $0<\l<1$,\cr
\cr
{(t\log{t})}^{-{1\over 3}} & if \quad $\l=1$, \cr
\cr
t^{-{1\over 2+\l}} & if \quad $\l>1$. \cr}
\eqnoi
$$

These results are, in fact, correct only for $1/2 \leq \l \leq 2$.
The upper bound follows from comparison of the random walk length,
$L_{\rm RW}\sim t^{1/2}$, with the drift length,
$L\sim t^{2/(2+\l)}$ when $\l>1$. For $\l>2$, $L_{\rm RW}\gg L$, so
a pair of charges can escape annihilation through a random walk, and
therefore the diffusion controls the dynamics. Thus, for $\l>2$,
the diffusion-controlled asymptotic behavior, $n\sim t^{-1/4}$,
is expected.  The lower bound stems from
the fact that an average force acting on any particle in the
{\it infinite-particles} 1D system becomes infinite for $\l \leq 1/2$.
It can be shown rigorously by deriving a Holtsmark-like [10]
force distribution (we leave it for Appendix A).
Here, we provide more qualitative arguments which take into account the
finiteness of the system. First, we note that in calculation of
the total force in (3), we implicitly assumed
that the system is perfectly ordered --- it consists of similar
domains of $A$ and $B$ particles; \ie, the total charge of the
first domain is equal, up to the sign, to the total charge
of the second domain, \etc\  In particular, it means that for a system
depicted in Fig.~1, the overall charge to the left of the test particle
$\tilde A$ is $-1$, and the overall charge to the right is zero.
However, this picture is a "mean-field" in spirit; hence, it can
lead to erroneous results for truly random systems. Fluctuations
in initial charge distribution in the system with $N$ particles produce
the net charge of the order $\sqrt N$ to the left and to the right of
the test particle.  Qualitatively, we can estimate the effect
of charge imbalance by putting $\sqrt N$ equidistant charges of one sign
to the right and the same amount of the opposite charges to the left.
(When we consider {\it finite} systems of charges we implicitly assume
that they satisfy the neutrality condition; generally,
the total net charge determines the long-time behavior).
The initial size of the system is $N$, so the distance between nearest
charges left in the system is $\sqrt N$; thus, the force $F_N$
due to the charge imbalance is
$$
F_N \sim {1\over N^{\l/2}}\sum_{j=1}^{\sqrt{N}}{1\over j^\l}\sim  N^{-\l+1/2}.
\eqnoi
$$
While $F_N\to 0$ as $N\to \infty$, this force does not affect the
dynamics of the model.  Thus, for $\l>1/2$ our previous estimate,
$F\sim L^{1/2-\l}$, gives the dominate contribution, and the
size-independent dynamics (7) emerges.  However, for $\l < 1/2$
the total force acting on a particle grows with system size even for
overall neutral systems; therefore, it should dominate over
the "regular" force (3) and control the dynamics of the system.
It seems reasonable to assume that at the early stages of time
evolution, when the distribution of the particles is still almost
random, the motion of domain edges is controlled by the force (8).
Repeating the steps used in deriving \back1, with $F \sim N^{1/2-\l}$
instead of \back2,  we obtain  for the density decay
$$
n(t)\sim N^{{2\l-1 \over 4}} t^{-{1\over 2}}.
\eqnoi
$$
However, this estimate may become inapplicable  on the later stages
of evolution. Indeed, the dynamics described by \last\ is extremely
fast since it is size-dependent, so the charge distribution that
emerges can be significantly different from the Poissonian; as a result,
our assumption about the type of randomness of the particle distribution
could become less and less appropriate.

\bigskip\bigskip
\chap{SIMULATION RESULTS}

To check our heuristic predictions, we have performed numerical simulations
for $\l=1,~0.75,~0.5,~0.25,$ and $0$ (the Coulomb case $\l=0$ turns out
to be special; it will be discussed separately in Appendix A).
Our system  initially consisted of $10000$ particles of each species
randomly distributed with concentration $1$.
First, the net force (1) is calculated for each particle.
We compute all the forces directly without applying any
multipole-like expansion that could be useful in many dimensions [12].
With the particle velocity equal to the total force, we
employ a simple Euler update procedure for each time step: $\D x_i
= F_i \D t$. The selection of time interval $\D t$ was merely
experimental. Since on the last stages of evolution simulations run very
fast (few particles are left), we,
unlike [7-8], keep $\D t$ constant during a run.
Finally the results are averaged over 10 runs.
The selection of boundary conditions does not seem to affect the results
of simulations of the two-component system with overall neutrality,
except for, maybe, its latest stages. We ran the simulations
for $\l=0.25$ with both periodic and open boundary conditions; the
results for concentration coincided within a statistical error.
Hence, we use open boundary conditions for all further simulations.

In Fig.~2 we plot, for various $\l$, the concentration of either species
\vs\ time. As we anticipated, for  $1/2 < \l$ our predictions for the
decay of concentration are in good agreement with the results of
simulation. Indeed,  the average slope of $\log{(n(t))}$ \vs\
$\log{t}$ plot for $\l=1$ is $-0.322$
[the heuristic argument gives ${(t\log{t})}^{-1/3}$],  and $-0.401$
(compared to $-2/5$) for $\l=3/4$.  Performing numerical simulations
with a twice smaller system ($5000$ pairs of particles),  we did not
find any significant system size dependence for these values of $\l$.
As for $\l \leq 1/2$, no power-law behavior can be observed for
density decay. However, at the early stages of evolution, local
exponents are  close to $-1/2$ as it follows from \last;
as time goes on, the density decay rate increases.

\bigskip\bigskip
\chap{DISCUSSION}

Considering the dynamics of domain interfaces and using simple
heuristic arguments, we have predicted the asymptotic density
decay in a two-species annihilation reaction system with long-range
interaction and overdamped motion. This approach can be
generalized on  similar systems in an arbitrary spatial dimension $d$.
Following the line of reasoning described in Sec.~II, one can obtain:
$$
n(t) \sim {1 \over \sqrt {L^d}} \sim \cases{t^{-{d\over 2-d+2\l}} & if
\quad $d/2\leq \l<d$,\cr
\cr
{(t\log{t})}^{-{d\over 2+d}} & if \quad $\l=d$, \cr
\cr
t^{-{d\over 2+\l}} & if \quad $\l>d$. \cr}
\eqnoi
$$
For $\l<d/2$, we expect size-dependent kinetics.  For $\l>2$,
as in the $1d$ case, let us compare diffusion and drift length scales,
$L_{\rm RW}\sim t^{1/2}$ and $L\sim  t^{2/(2+\l)}$, and conclude that
the random walk dominates the drift; hence, the
diffusion-controlled behavior, $n\sim t^{-d/4}$, is expected.
Thus, the regime described by the lower
line of \last\ does not appear for $d\ge 2$. Note that for truly
Coulomb systems, $\l=d-1$, the scaling prediction of \last\ is
$n\sim t^{-1}$; \ie, the classical kinetic law.  This is expected to
arise when $d/2\leq \l=d-1\leq 2$, \ie, $2\leq d\leq 3$. For $d<2$,
size-dependent kinetics is anticipated; while for $d>3$, the Coulomb
interaction becomes irrelevant and the diffusion-controlled behavior
emerges.  Another interesting example is the system of $D$-dimensional
Coulomb charges confined to the hypersurface; \ie, $\l=d=D-1$. In this
case, \last\ predicts logarithmic corrections to the power law behavior.

However, it is not clear whether the concept of unpenetrable
(untransparent) domains with continuous boundaries is still applicable
for $d>1$. Competition between screening, which makes long-range
interaction effectively short-range, and annihilation may also
significantly affect the behavior of the system. We attempted to
simulate a $2d$ system on a lattice. Either because of the large effective
diffusion, which is inevitably introduced by the discrete nature
of the lattice model, or for some deeper reason,
we were unable to observe any scaling-like behavior.
Since many-dimensional continuous many-body simulations are still
computationally challenging, we leave this problem for the future.

More can probably be done with one-dimensional systems as well.
One can try to find exact results for some specific values of the force
exponent $\l$. For the Coulomb system, $\l=0$, we indeed succeeded in
finding some properties analytically (see Appendix A and Ref.~[11]), but
we still could not find a complete solution. Another extreme case,
$\l\to\infty$, is also theoretically challenging. (In fact, the diffusion
determines the dynamics for $\l>4$ so one should consider a {\it strictly}
noiseless system).  The dynamics in the $\l\to\infty$ limit is
extremal: One picks the  pair of nearest neighbors that are
{\it closest} to each other and removes it if the charges
are dissimilar, or recedes them if the charges are similar.
The receding is stopped when the distance reaches the second minimal
intercharge distance.  Then, if this second closest pair contains
dissimilar charges, it is removed while the first pair continues receding;
if the second pair is also the same-species, both pairs recede.
If the initial sequence is alternating
as it takes place with domain walls in the quench process, one always
removes the closest pairs, and the domain-size distribution function
approaches the scaling form. This model turns out to be completely
solvable [13-15,8]. It would be very interesting to study a more
complex version of the extremal dynamics
that arises from the present two-species annihilation process.

Finally, we discuss a model where motion of the particles is
ballistic; \ie, it is described by Newton's laws.  It is clear enough
that the only change one needs to make in the above approach is to put
$d^2 L/dt^2$ instead of $dL/dt$ into the left hand side of (6).
Assuming that initial velocities are irrelevant, we obtain for the
concentration
$$
n(t) \sim  \cases{N^{{2\l-1 \over 4}} t^{-1}& if \quad $\l<1/2$,\cr
\cr
t^{-{2\over 1+2\l}} & if \quad $1/2<\l<1$,\cr
\cr
{(t\log{t})}^{-{2\over 3}} & if \quad $\l=1$, \cr
\cr
t^{-{2\over 2+\l}} & if \quad $1<\l<2$, \cr
t^{-{1\over 2}} & if \quad $\l\geq 2$. \cr}
\eqnoi
$$
When $\l\geq 2$, the inertia dominates the drift and, therefore, the
ballistic-controlled asymptotic behavior [16], given by the last line of
\last, follows.

Numerical simulations performed for $\l=0.75$ showed that
the concentration decays as $t^{-0.79}$ compared to $t^{-4/5}$ as it
follows from \last.

\bigskip\bigskip\bigskip
\centerline{\bf Acknowledgments}\medskip
We are thankful to S.~Redner for numerous discussions.  We gratefully
acknowledge ARO grant \#DAAH04-93-G-0021 for partial support
of this research.

\vfill\eject
\centerline{\bf APPENDIX A: $\l=0$}\bigskip

The $\l=0$ case, corresponding to the "real" $1d$  Coulomb force has several
peculiar features. Since the forces
between particles do not depend on the distance, and
particles disappear in pairs, the net force acting on a particle is
constant throughout its life. It means, that velocity of any
particle is constant and equal to the difference between total charge to
the left and total charge to the right of it multiplied by a charge of
the particle. Taking into account that initial distribution of particles
is Poissonian, one can easily describe the behavior of the system if the
probability distribution for "charge imbalance" were known.
To get an insight about this charge imbalance distribution, we look at our
configuration of charges as on a $1d$ random walk (RW) (Fig.3).
Step up corresponds to a positive charge, step down -- to the negative;
since the system is overall neutral, the RW returns to the origin after
$2N$ (the size of the system) steps.
The net force acting on a particle is equal to the "height" $h$,
positive or negative, of the corresponding point in the RW
picture. The joint distribution function $W_{2N}(h,2L)$ tells us
how many segments (loops in the RW terminology) of the length $2L$,
starting and ending at the "height" $h$ from the origin,  exist in a
RW coming to the origin (not necessarily for the first
time) after $2N$ steps. Knowing this function, one can readily calculate
a life expectation time for each particle, and therefore the concentration
decay rate:
$$
{dn\over dt}=-{1\over \tilde L} \sum_{L=0}^N {\sum_{h=0}^{N-L}
{W_{2N}(h,2L)P_{2L}[2t(2h+1)]}}.
\eqno(A1)
$$
Here $\tilde L$ is initial system length and $P_j(x)=x^je^{-x}/j!$
the Poisson distribution function. So far we have been able to determine
another function, $\tilde W_{2N}(2L)$, which gives the probability
that the {\it maximum} length of "zero-height" segment in a system
with $2N$ charges is $2L$ [11]. This maximum length segment determines
the lifetime of the whole system, which is shown to be proportional
to its size $2N$. Moreover, the life time distribution function has
a remarkably rich structure (an infinite set of singularities, \etc;
see [11]).

Although we were not able to find the exact expression for the decay of
concentration in this case, numerical simulations of this problem prove
to be very simple. Instead of running a molecular dynamics algorithm,
it is sufficient to calculate all the net forces once and find an
annihilation partner for each particle; after that, we know the lifetimes for
all the particles in the system. It enabled us to check an accuracy of
our molecular dynamics simulation for our standard ($10^4$ pairs) system
size and also to study the systems with up to $10^5$ particles of each
species. Even for these relatively big systems, we were unable to find any
power-law behavior; the function that fits best our simulation results
looks like $n(t)\sim \exp{(-0.05 \ln^4 t)}$.

\bigskip\bigskip\centerline{\bf APPENDIX B:}
\centerline{\bf TYPICAL FORCE IN A SYSTEM OF CHARGED PARTICLES}\bigskip

A problem of distribution of a typical force in a system of particles
with Coulomb interaction was first studied by Holtsmark [10].
Using the same approach, we will show, that for arbitrary dimensionality and
$r^{-\l}$ interaction, the typical (or mean) force is
finite in an infinite system only for some range of the
force constants $\l$, specifically for $\l>d/2$.
We start from an expression for force distribution function $W(F)$,
which gives the probability that the force, acting on a "test particle"
which we will put at the origin, is equal to $F$:
$$
W(\vec F)=\langle \d(\vec F-\sum_{j=1}^N {\vec f(\vec r_j))\rangle_{av}},
\eqno(B1)
$$
where $\vec f(\vec r_j)=\vec r_j/r_j^{\l+1}$ is the force exerted by
a \jth\ particle on the test one.
Since we assume that spatial distribution of particles is
random and independent, it is sufficient to consider one-component
system.
Replacing $\d$-function by auxiliary integration over $d \vec k$
we obtain:
$$
W(\vec F)={1\over (2\p)^d} \int {\exp{(i\vec k\cdot \vec F)}
S(\vec k)d\vec k}
\eqno(B2)
$$
with
$$
S(\vec k)=\int\ldots\int
\exp\left(-\sum_{j=1}^N {i\vec k\cdot\vec f(\vec r_j)}\right)
\prod_{j=1}^N {d\vec r_j \over V}\equiv
\left({1\over V}\int e^{-i\vec k\cdot\vec f}d\vec r\right)^N.
\eqno(B3)
$$
Here $\vec f=\vec r/r^{\l+1}$, $N$ is the number of particles in the
system,  and $V$ is the volume.  Rewriting (B3) in the form
$$
S(\vec k)=\left[1-{1\over V}\int
\left(1-e^{-i\vec k\cdot\vec f}\right)d\vec r\right]^N
\eqno(B4)
$$
and taking the thermodynamic limit,  $N\rightarrow \infty$
and $V\rightarrow \infty$ where $N/V=n$ is kept fixed, yields
$$
S(\vec k)=\exp\left(-n\int\left[1-e^{-i\vec k\cdot
\vec r/r^{\l+1}}\right]
d\vec r\right).
\eqno(B5)
$$

After expanding the exponent in the integrand for large $r$ and performing
angular integration (which eliminates all odd-order terms) one finds
that for system size $R \rightarrow \infty$, the integral in the
right-hand side of Eq.~(B5) converges only if $\l>d/2$ while for $\l<d/2$
the integral diverges with the system size. More precisely,
a considerable but straightforward computation yields
$$
S(\vec k)\equiv S(k)=\exp\left[-n\Omega_d A(d,\l)k^{d\over \l}\right]
\eqno(B6)
$$
for $\l>d/2$. In Eq.~(B6), $\Omega_d=2\pi^{d/2}/\G(d/2)$ denotes the surface
area of the unit sphere in $d$ dimensions and $A(d,\l)$ is the shorthand
notation for the integral
$$
A(d,\l)=\int_0^\infty {dz\over \l z^{{d\over \l}+1}}
\left[1-\G\Bigl({d\over 2}\Bigr)\Bigl({2\over z}\Bigr)^{{d\over 2}-1}
J_{{d\over 2}-1}(z)\right],
$$
with $\G(z)$ being the Euler gamma function, and $J_\nu(z)$
the Bessel function.
For $\l<d/2$ the integral in the left-hand side of Eq.~(B5) grows
as ${2\Omega_d\over d(d-2\l)}R^{d-2\l}k^2$.
Returning back to Eq.~(B2) we see that in the case of
($\l\leq d/2$) the net force is given by
$$
W(\vec F)={1\over (2\p)^d} \int \exp\left(i\vec k\cdot \vec F
-{2\Omega_d\over d(d-2\l)}R^{d-2\l}k^2\right)d\vec k.
\eqno(B7)
$$
In the limit $R\to \infty$, we compute the integral in Eq.~(B7)
asymptotically to find
$$
W(\vec F)\sim \left[(d-2\l)R^{2\l-d}F\right]^{d-1}
\exp\left[-{\rm const}~(d-2\l)R^{2\l-d}F^2\right].
\eqno(B8)
$$
Eqs.~(B7) and (B8) are valid for $\l<d/2$. For $\l=d/2$,
$R^{d-2\l}/(d-2\l)$ should be replaced by $\log R$.
For $d=1$ and $R \propto N$, the net force has the Gaussian
distribution, $W(\vec F)\propto (1-2\l)^{1/2}N^{\l-1/2}
\exp\left[-(1-2\l)N^{2\l-1}F^2\right]$,
and therefore the typical force grows with size as $N^{1/2-\l}$
in agreement with qualitative results of Sec.~III.  Note also that
for $\l=1/2$ the typical force still grow with system size,
$F\sim \sqrt{\log N}$, and hence the density decay of the form
$n(t)\sim (\log N)^{-1/4}t^{-1/2}$ is expected.

\vfill\eject

{
\parindent=0.2in
\centerline{\bf References}\medskip

\refi For a recent review, see ~S.~Redner and F.~Leyvraz,
      in {\sl Fractals in Science}, eds. A.~Bunde and S.~Havlin
      (Springer-Verlag, Berlin, 1994).

\refi T.~Ohtsuki, \pla 106 224 84 .

\refi V.~Kuzovkov and E.~Kotomin, \jsp 72 127 1993 .

\refi M.~Mandello and N.~Goldenfeld, \pra 42 5865 1990 .

\refi W.~G.~Jang, V.~V.~Ginzburg, C.~D.~Muzny, and N.~A.~Clark,
      \pre 51 411 1995 .

\refi F.~J.~Dyson, \cmp 12 91 1969 .

\refi B.~P.~Lee and J.~L.~Cardy, \pre 48 2452 1993 .

\refi A.~D.~Rutenberg and A.~J.~Bray, \pre 50 1900 1994 .

\refi T.~Ohta and H.~Hayakawa, \pA 204 482 1994 .

\refi See, \eg, S.~Chandrasekhar, \rmp 15 1 1943 .

\refi L.~Frachebourg, I.~Ispolatov and P.~L.~Krapivsky, unpublished.

\refi L.~Greengard, {\sl The Rapid Evaluation of Potential Fields in
      Particle Systems}, (MIT Press, Cambridge, 1988).

\refi T.~Nagai and K.~Kawasaki, \pA 134 483 1986 ;
      K.~Kawasaki, A.~Ogawa, and T.~Nagai, \pB 149 97 1988 .

\refi A.~J.~Bray, B.~Derrida, and C.~Godreche, \eul 27 175 1994 ;
      A.~J.~Bray and B.~Derrida, \pre 51 1633 1995 .

\refi S.~N.~Majumdar and D.~A.~Huse, \pre 52 270 1995 .

\refi Y.~Elskens and H.~L.~Frisch, \pra 31 3812 1985 .
}

\vfill\eject
\centerline{\bf Figure Captions}\bigskip

\rfigi  Schematic illustration of domain structure. Typical domain has
length $L$ and consists of $M=6$ particles, average distance between
particles is $R=L/N$.

\rfigi  Plot of concentration $n(t)$ \vs\ time on a double logarithmic scale
for $\l=1 (\diamond), \l=3/4 (\bigtriangleup), \l=1/2 (\bigtriangledown)
\l=1/4 (\bigcirc)$, and $\l=0 (\star)$.

\rfigi  RW representation of a two species neutral system. Number of
particles of each species $N=17$. A neutral segment of the length
$2L=16$ is shown having $h=2$ uncompensated charges to the left and to
the right of it.

\vfill\eject\bye